\documentclass{sig-alternate}
\usepackage[T1]{fontenc}
\usepackage{graphicx}
\usepackage{epsfig}
\usepackage{color}
\usepackage{latexsym}
\usepackage{amssymb}
\usepackage{amsmath}
\usepackage{xspace}
\usepackage{url}
\usepackage{pslatex}
\usepackage{ifpdf}
\usepackage{cite}
\usepackage{balance}
\usepackage{fixacm}

\ifpdf
\fi

\def\more-auths{%
\end{tabular}
\begin{tabular}{c}}
\DeclareSymbolFont{msbm}{U}{msb}{m}{n}
\DeclareMathSymbol{\C}{\mathalpha}{msbm}{'103}
\DeclareMathSymbol{\R}{\mathalpha}{msbm}{'122}
\DeclareMathSymbol{\Q}{\mathalpha}{msbm}{'121}
\DeclareMathSymbol{\Z}{\mathalpha}{msbm}{'132}
\DeclareMathSymbol{\N}{\mathalpha}{msbm}{'116}
\DeclareMathSymbol{\K}{\mathalpha}{msbm}{'113}

\newtheorem{lemma}{Lemma}

\newtheorem{theorem}[lemma]{Theorem}

\renewcommand{\qed}{\hfill{\ensuremath{\Box}}}
\newcommand{\eps}{\varepsilon}

\newcommand{\abs}[1]{\mathopen| #1 \mathclose|}

\newcommand{\bi}{\begin{itemize}}
\newcommand{\ei}{\end  {itemize}}
\newcommand{\bt}{\begin{tabbing}}
\newcommand{\et}{\end  {tabbing}}
\newcommand{\be}{\begin{enumerate}}
\newcommand{\ee}{\end  {enumerate}}

\renewcommand{\leq}{\leqslant}  %needs \usepackage{amssymb}
\renewcommand{\geq}{\geqslant}

\newcommand{\atan}{\tan^{-1}}

\makeatletter
\def\begin@lgo{\begin{minipage}{1in}\begin{tabbing}
        \quad\=\qquad\=\qquad\=\qquad\=\qquad\=\qquad\=\qquad\=\kill}
\def\end@lgo{\end{tabbing}\end{minipage}}
\newenvironment{algorithm}{\begin{tabular}{|l|}\hline\begin@lgo}{\end@lgo\\\hline\end{tabular}}
\makeatother

\makeatletter
\@ifundefined{abovecaptionskip}{\newlength\abovecaptionskip}
\long\def\@makecaption#1#2{
   \vskip \abovecaptionskip
   \setbox\@tempboxa\hbox{{\sf\footnotesize \textbf{#1.} #2}}
   \ifdim \wd\@tempboxa >\hsize         % IF longer than one line:
       {\sf\footnotesize \textbf{#1.} #2\par}% THEN set as ordinary paragraph.
     \else                              %   ELSE  center.
       \hbox to\hsize{\hfil\box\@tempboxa\hfil}
   \fi}
\makeatother

\begin{document}

\conferenceinfo{SCG'06,} {June 5--7, 2006, Sedona, Arizona, USA.}
\CopyrightYear{2006}
\crdata{1-59593-340-9/06/0006}

\title{Minimum-Cost Coverage of Point Sets by Disks{\huge \footnotemark[1]}}

\numberofauthors{3} % NOTE! There are still '3' authors (Rous, Richards and Unson).
                    % The other authors (Siedun, Rodkin and Simon), although they may format underneath,
                    % are not in the author count -- they are merely being 'patched in'.

\author{
\alignauthor Helmut Alt\\
%\affaddr{Institute of Computer Science}\\
\affaddr{Freie Universit{\"a}t Berlin}\\
\affaddr{D-14195 Berlin, Germany}\\
%\email{alt@inf.fu-berlin.de}
%
\alignauthor Esther M. Arkin\\
%\affaddr{Department of Applied Mathematics and Statistics}\\
\affaddr{Stony Brook University}\\
\affaddr{Stony Brook, NY 11794, USA}  %, USA
%\email{estie@ams.sunysb.edu}
%
\alignauthor Herv\'e Br\"onnimann\\
%\affaddr{Computer and Information Science}\\
\affaddr{Polytechnic University}\\
\affaddr{Brooklyn, NY 11201, USA}  %, USA
%\email{hbr@poly.edu}
%
\vspace{3mm}% optional
\more-auths
\alignauthor Jeff Erickson\\
%\affaddr{Department of Computer Science}\\
\affaddr{University of Illinois}\\
\affaddr{Urbana, IL 61801, USA}  %, USA
%\email{jeffe@cs.uiuc.edu}
%
\alignauthor S\'andor P.\ Fekete\\
%\affaddr{Department of Mathematical Optimization}\\
\affaddr{TU Braunschweig}\\ %% of Technology
\affaddr{D-38106 Braunschweig, Germany}\\
%\email{s.fekete@tu-bs.de}
%
\alignauthor Christian Knauer \\
%\affaddr{Institute of Computer Science}\\
\affaddr{Freie Universit{\"a}t Berlin}\\
\affaddr{D-14195 Berlin, Germany}\\
%\email{knauer@inf.fu-berlin.de}
%
\vspace{3mm}% optional
\more-auths
\alignauthor Jonathan Lenchner\\
\affaddr{IBM T. J.\ Watson Research} \\ % Center
\affaddr{Yorktown Heights, NY 10598, USA}  %, USA
%\email{lenchner@us.ibm.com}
%
\alignauthor Joseph S. B. Mitchell\\
%\affaddr{Department of Applied Mathematics and Statistics}\\
\affaddr{Stony Brook University}\\
\affaddr{Stony Brook, NY 11794, USA}  %, USA
%\email{jsbm@ams.sunysb.edu}
%
\alignauthor Kim Whittlesey\\
%\affaddr{Department of Mathematics}\\
\affaddr{University of Illinois}\\
\affaddr{Urbana, IL 61801, USA}\\
%\email{kwhittle@math.uiuc.edu}
}

\date{}

\maketitle

\renewcommand{\thefootnote}{\fnsymbol{footnote}}

\footnotetext[1]{
E. Arkin is partially supported by grants
from the National Science Foundation (CCR-0098172,
CCF-0431030).
H.~Br\"onnimann and J.~Lenchner are partially supported by a grant
from the National Science Foundation (Career grant CCR-0133599).
J.~Mitchell is partially supported by grants from
the National Science Foundation (CCR-0098172,
ACI-0328930, CCF-0431030, CCF-0528209),
the U.S.-Israel Binational Science Foundation (2000160),
Metron Aviation, and NASA Ames (NAG2-1620).}

\renewcommand{\thefootnote}{\arabic{footnote}}

\begin{abstract}
  We consider a class of geometric facility location problems in which
  the goal is to determine a set $X$ of disks given by their centers
  ($t_j$) and radii ($r_j$) that cover a given set of demand points
  $Y\subset \R^2$ at the smallest possible cost.
  We consider cost functions of the form $\sum_j f(r_j)$, where
  $f(r)=r^\alpha$ is the cost of transmission to radius $r$.  Special
  cases arise for $\alpha=1$ (sum of radii) and $\alpha=2$ (total
  area); power consumption models in wireless network design often use
  an exponent $\alpha>2$.  Different scenarios arise according to
  possible restrictions on the transmission centers $t_j$, which may
  be constrained to belong to a given discrete set or to lie on a line,
  etc.

  We obtain several new results, including (a) exact and approximation
  algorithms for selecting transmission points $t_j$ on a given line
  in order to cover demand points $Y\subset \R^2$; (b) approximation
  algorithms (and an algebraic intractability result) for selecting an
  optimal line on which to place transmission points to cover $Y$;
  (c) a proof of NP-hardness for a discrete set of transmission points
  in $\R^2$ and any fixed $\alpha>1$; and (d) a polynomial-time
  approximation scheme for the problem of
  computing a \emph{minimum cost covering tour} (MCCT), in which the
  total cost is a linear combination of the transmission cost for the
  set of disks and the {\em length} of a tour/path that connects the
  centers of the disks.
\end{abstract}

%%%%%%%%%%%%%%%%%%%%%%%%%%%%%%%%%%%%%%%%%%%%%%%%%%%%%%%%%%%%%%%%%%%%%%%

\vspace{-1mm}

\vspace{-3mm }

%\vspace{-3mm}

%
%{\bf ACM Classification:} F.2.2 Nonnumerical Algorithms and Problems.
%%, G.2.1 Combinatorics, G.2.2 Graph Theory, G.1.6 Optimization

%{\bf AMS Classification:} 68Q25, 68U05, 90C27.

%{\bf Keywords:} Covering problems, tour problems,
%geometric optimization, complexity, approximation.

\setcounter{page}{1}

%%%%%%%%%%%%%%%%%%%%%%%%%%%%%%%%%%%%%%%%%%%%%%%%%%%%%%%%%%%%%%%%%%
\section{Introduction}
\label{sec:intro}
%%%%%%%%%%%%%%%%%%%%%%%%%%%%%%%%%%%%%%%%%%%%%%%%%%%%%%%%%%%%%%%%%%

\newcommand{\mypara}[1]{{\smallskip\noindent {\bf #1}}\quad}

\mypara{The problem.}
We study a geometric optimization problem that arises in wireless
network design, as well as in robotics and various facility location
problems. The task is to select a number of locations $t_j$ for the
base station antennas (\emph{servers}), and assign a transmission
range $r_j$ to each $t_j$, in order that each $p_i\in Y$ for a given
set $Y=\{p_1,\ldots,p_n\}$ of $n$ demand points (\emph{clients}) is covered.  We say that
client $p_i$ is covered if and only if $p_i$ is within range of some transmission
point $t_{j_i}$, i.e., $d(t_{j_i},p_i)\leq r_{j_i}$. The resulting
cost per server is some known function $f$, such as $f(r)=r^\alpha$.
The goal is to minimize the total cost, $\sum_j f(r_j)$, over all
placements of at most $k$ servers that cover the set $Y$ of clients.

In the context of modeling the energy required for wireless
transmission, it is common to assume a superlinear ($\alpha>1$)
dependence of the cost on the radius; in fact, physically accurate
simulation often requires superquadratic dependence ($\alpha>2$).  A
quadratic dependence ($\alpha=2$) models the total area of the served
region, an objective arising in some applications.  A linear
dependence ($\alpha=1$) is sometimes assumed, as in Lev-Tov and
Peleg\cite{lp-ptasb-05}, who study the base station coverage problem,
minimizing the sum of radii.  The linear case is important to study
not only in order to simplify the problem and gain insight into the
general problem, but also to address those settings in which the
linear cost model naturally arises\cite{cps-opapr-00,210191}.  For
example, the model may be appropriate for a system with a narrow-angle
beam whose direction can either rotate continuously or adapt to the
needs of the network.  Another motivation for us comes from robotics,
in which a robot is to map or scan an environment with a laser
scanner\cite{fkn-osar-05,fkn-sar-04}: For a fixed spatial resolution
of the desired map, the time it takes to scan a circle corresponds to
the number of points on the perimeter, i.e., is proportional to the
radius.

Our problem is a type of clustering problem, recently named
\emph{min-size $k$-clus\-ter\-ing} by Bil\`o et al.\cite{ESA05}.
Clustering problems tend to be NP-hard, so most efforts, including
ours, are aimed at devising an approximation algorithm or
a polynomial-time approximation scheme (PTAS).

We also introduce a new problem, which we call \emph{minimum cost
  covering tour} (MCCT), in which we combine the problem of finding a
short tour and placing covering disks centered along it.  The
objective is to minimize a linear combination of the tour length and
the transmission/covering costs.  The problem arises in the autonomous
robot scanning problem\cite{fkn-osar-05,fkn-sar-04}, where the
covering cost is linear in the radii of the disks, and the overall
objective is to minimize the total time of acquisition (a linear
combination of distance travelled and sum of scan radii).  Another
motivation is the distribution of a valuable or sensitive resource:
There is a trade-off between the cost of broadcasting from a central
location (thus wasting transmission or risking interception) and the
cost of travelling to broadcast more locally, thereby reducing
broadcast costs but incurring travel costs.

\mypara{Location Constraints.}
In the absence of constraints on the server locations, it may be
optimal to place one server at each demand point.  Thus, we generally
set an upper bound, $k$, on the number of servers, or we restrict the
possible locations of the servers.  Here, we consider two cases of
location constraints:

(1) Servers are restricted to lie in a discrete
set $\{t_1,\ldots,t_m\}$; or

(2) Servers are constrained to lie on a
line (which may be fully specified, or may be selected by the
optimization).

\mypara{Our results.}
We provide a number of new results, some improving previous
work, some giving the first results of their kind.

In the discrete case studied by Lev-Tov and
Peleg\cite{lp-ptasb-05}, and Bil\'o et al.\cite{ESA05}, we
give improved results.
For the discrete 1D problem where $Y\subseteq\mathbb{R}$, we improve their 4-approximation to a linear-time 3-approximation by using a ``Closest
Center with Growth'' (CCG) algorithm, and, as an alternative to the
previous $O((n+m)^3)$ algorithm\cite{lp-ptasb-05},
we give a near-linear-time
2-approximation that uses a ``Greedy Growth'' (GG) algorithm.
Unfortunately, we cannot extend our proofs to the 1$\frac12$D problem.
Intuitively, greedy growth works as follows: start with a disk with
center at each server, each disk of radius zero; among all clients, find
one that requires the least radial disk growth to capture it; repeat
until all clients are covered. Note that for $\alpha\geq 2$ the 2D variants of the
problem are already proved to be NP-Hard and to have a
PTAS\cite{ESA05}.

In the general 2D case with clients $Y\subset \R^2$, we strengthen
the hardness result of Bil{\'o} et al.~\cite{ESA05} by showing
that the discrete problem is already hard for any superlinear cost
function, i.e., $f(r)=r^{\alpha}$ with $\alpha>1$. Furthermore,
we generalize the min-size clustering problem in two new directions.  On the one
hand, we consider less restrictive server placement policies.  For
instance, if we only restrict the servers to lie on a given fixed
line, we give a dynamic programming algorithm that solves the
problem exactly, in time $O(n^2\log n)$ for any $L_p$ metric in
the linear cost case, and in time $O(n^4\log n)$ in the case of
superlinear non-decreasing cost functions.  For simple
approximations, our algorithm ``Square Greedy'' (SG) gives in time
$O(n \log n)$ a $3$-approximation to the square covering problem
with any linear or superlinear cost function.  A small variation,
``Square Greedy with Growth'' (SGG), gives a 2-approximation for a
linear cost function, also in time $O(n\log n)$.  The results are
also valid for covering by $L_p$ disks for any~$p$, but with
correspondingly coarser approximation factors.

A practical example in which
servers are restricted to lie along a line is that of
a highway that cuts through a piece of land, and the server
locations are restricted to lie along the highway.  The line location
problem arises when one not only needs to locate the servers, but also
needs to select an optimal corridor for the placement of the highway.
Other relevant examples may include devices powered by a microwave or
laser beam lining up along the beam.

If the servers are restricted to lie on a horizontal line, but the
location of this line may be chosen freely, then we show that the
exact optimal position (with $\alpha=1$) is not computable by
radicals, using an approach similar to that of
Bajaj\cite{b-pgans-86,b-adgop-88} in addressing the unsolvability
of the Fermat-Weber problem.  On the positive side, we give a
fully polynomial-time approximation scheme (FPTAS) requiring time
$O((n^3/\eps) \log n)$ if $\alpha=1$ and time $O((n^5/\eps) \log
n)$ if $\alpha>1$.

For servers on an unrestricted line, of any slope, and $\alpha=1$, we
give $O(1)$-approximations (4-approximation in $O(n^4\log n)$ time, or
$8\sqrt{2}$-approximation in $O(n^3\log n)$ time) and an FPTAS
requiring time $O((n^5/\eps^2)\log n)$.

We give the first algorithmic results for the new problem, minimum
cost covering tour (MCCT), which we introduce.  Given a set
$Y\subseteq\mathbb{R}^2$ of $n$ clients, our goal is to
determine a polygonal tour $T$ and
a set $X$ of $k$ disks of radii $r_j$ centered on $T$
that cover $Y$ while minimizing
the cost $\mathrm{length}(T)+C\sum r_i^\alpha$.
Our results are for $\alpha=1$.
 The ratio $C$ represents the relative cost of touring versus transmitting.
We show that MCCT is NP-hard if $C$ is part of
the input. At one extreme, if $C$ is small
 then the optimum solution is a single server placed at the
circumcenter of $Y$ (we can show this to be the case for $C\leq 4$).
At the other extreme (if $C$ very large), the
optimum solution is a TSP among the clients.  For any fixed value of
$C>4$, we present a PTAS for MCCT, based on a novel extension of the
$m$-guillotine methods of~\cite{m-gsaps-99}.

\mypara{Related work.}
There is a vast family of clustering problems,
among which are the \emph{$k$-center} problem in which one
minimizes $\max_j r_j$, the \emph{$k$-median} problem in which one
minimizes $\sum_i d(p_i,t_{j_i})$, and the \emph{$k$-clustering}
problem in which one minimizes the maximum over all clusters of
the sum of pairwise distances between points in that cluster. For
the geometric instances of these related clustering problems,
refer to the survey by Agarwal and Sharir\cite{299918}. When $k$
is fixed, the optimal solution can be found in time $O(n^k)$ using
brute force. In the plane, one of the only results for the
min-size clustering problem is a small improvement for $k=2$ by
Hershberger\cite{160550}, in subquadratic time $O(n^2/\log\log
n)$. Approximation algorithms and schemes have been proposed,
particularly for geometric instances of these problems (e.g.,
\cite{Arora:1998tw}). Clustering for minimizing the sum of radii
was studied for points in metric spaces by Charikar and
Panigrahy\cite{985670}, who present an $O(1)$-approximation
algorithm using at most $k$ clusters.

For the linear-cost model ($\alpha=1$), our problem has been
considered recently by Lev-Tov and Peleg\cite{lp-ptasb-05}
who give an $O((n+m)^3)$ algorithm when the clients and servers all
lie on a given line (the 1D problem), and a linear-time
4-approximation in that case.  They also give a PTAS for the
two-dimensional case when the clients and servers can lie anywhere in
the plane. Bil\`o et al.\cite{ESA05} show that the problem is NP-hard
in the plane for the case $f(r)=r^\alpha$, $\alpha\geq 2$, either when
the sets $X$ and $Y$ are given and $k$ is left unspecified ($k=n$), or
when $k$ is fixed but then $X=Y$.  They give a PTAS for the linear
cost case ($\alpha=1$) and a slightly more involved PTAS for
a more general problem in which the cost function is superlinear,
there are fixed additive costs associated with each transmission server and
there is a bound $k$ on the number of servers.

There are many problems dealing with covering a set of clients by
disks of {\em given} radius. Hochbaum and Maass\cite{hm-ascpp-85} give
a PTAS for covering with a minimum number of disks of fixed radius,
where the disk centers can be taken anywhere in the plane.  They
introduce a ``grid-shifting technique,'' which is used and extended by
Erlebach~et~al.\cite{ejs-ptasg-05}.  Lev-Tov and
Peleg\cite{lp-ptasb-05} and Bil\`o et al.\cite{ESA05} extend the
method further in obtaining their PTAS results for the discrete
version of our problem.

When a discrete set $X$ of potential server locations is given,
Gonzalez\cite{Gonzalez:1991bb} addresses the problem of maximizing
the number of covered clients while minimizing the number of
servers supplying them, and he gives a PTAS for such problems with
constraints such as bounded distance between any two chosen
servers. In\cite{bg-aoscf-95}, a polynomial-time constant
approximation is obtained for choosing a subset of minimum size
that covers a set of points among a set of candidate disks (the
radii can be different but the candidate disks must be given).

The closest work to our combined tour/transmission cost (MCCT) is the
work on covering tours: the ``lawn mower'' problem\cite{Arkin:2000bd},
and the TSP with neighborhoods\cite{ah-aagcs-94,dm-aatsp-03}, each of
which has been shown to be NP-hard and has been solved with various
approximation algorithms. In contrast to the MCCT we study, the radius
of the ``mower'' or the radius of the neighborhoods to be visited is
specified in advance.

%%%%%%%%%%%%%%%%%%%%%%%%%%%%%%%%%%%%%%%%%%%%%%%%%%%%%%%%%%%%%%%%%%%%%%%
\section{Scenario (1): Server Locations Restricted to a Discrete Set}
\label{sec:scenario1a}
%%%%%%%%%%%%%%%%%%%%%%%%%%%%%%%%%%%%%%%%%%%%%%%%%%%%%%%%%%%%%%%%%%%%%%%

\subsection{The one-dimensional discrete problem with linear cost}

Consider the case of $m$ fixed
server locations $X=\{t_1,...,t_m\}$, $n$ client locations
$Y=\{p_1,...,p_n\}$, and a linear ($\alpha=1$) cost function, with
clients and servers all located along a fixed line.
Without loss of generality, we may assume that $X$ and $Y$ are sorted in the same direction, at an extra
cost of $O((n+m)\log(n+m))$.
Lev-Tov and Peleg\cite{lp-ptasb-05} give an $O((n+m)^3)$ dynamic programming algorithm for finding an exact
solution.  Bil\`o et al.\cite{ESA05} show that the problem is solvable
in polynomial time for any value of $\alpha$ by reducing it to an
integer linear program with a totally unimodular constraint matrix.
The complexities of these algorithms, while polynomial, is high.
Lev-Tov and Peleg also give a simple ``closest center'' algorithm (CC)
that gives a linear-time $4$-approximation.
We improve to a $3$-approximation in linear time, and a 2-approximation in $O(m+n\log m)$ time.

We now describe an algorithm which also runs in linear time,
but achieves an approximation factor of $3$.

\smallskip
\noindent \textbf{Closest Center with Growth (CCG) Algorithm}:
Process the clients $\{p_1,...,p_n\}$ from left to right keeping
track of the rightmost extending disk.  Let $\omega_R$ denote the
rightmost point of the rightmost extending disk, and let $R$
denote the radius of this disk. (In fact the rightmost extending
disk will always be the last disk placed.)  If $\omega_R$ is equal
to, or to the right of the next client processed, $p_i$, then
$p_i$ is already covered so ignore it and proceed to the next
client.  If $p_i$ is not yet covered, consider the distance of
$p_i$ to $\omega_R$ compared with the distance of $p_i$ to its
closest center $\hat{t_i}$.  If the distance of $p_i$ to
$\omega_R$ is less than or equal to the distance of $p_i$ to its
closest center $\hat{t_i}$, then grow the rightmost extending disk
just enough to capture $p_i$.  Otherwise use the disk centered at
$\hat{t_i}$ of radius $|p_i - \hat{t_i}|$ to cover $p_i$.

\begin{lemma} For $\alpha=1$, CCG yields a $3$-approximation to OPT in $O(n+m)$ time.
\end{lemma}

The proof is similar to that of the next lemma, and omitted in this version.

If we consider a single disk $D$ with clients $p_L$ and $p_R$ on
the left and right edges of $D$, associated centers $\hat{x}_L$,
$\hat{x}_R$ at distances respectively radius($D$)$- \epsilon$ to
the left and radius($D$)$- \epsilon$ to the right, along with a
dense set of clients in the left hand half of $D$ we see that $3$
is the best possible constant for CCG.

Finally we offer an algorithm that achieves a $2$-approximation
but runs in time $O(m+n\log{m})$.

\smallskip
\noindent \textbf{Greedy Growth (GG) Algorithm}: Start with a disk
with center at each server all of radius zero. Now, amongst all
clients, find the one which requires the least radial disk growth
to capture it.  Repeat until all clients are covered. An efficient
implementation uses a priority queue to determine the client that
should be captured next. One can set up the priority queue in
$O(m)$ time. Note that the priority queue will never have more
than $2m$ elements, and that each $p_i$ eventually gets captured,
either from the right or from the left. Each capture can be done
in time $O(\log{m})$ for a total running time of $O(m+n\log{m})$.

\begin{lemma} \label{l:2-approx}
For $\alpha=1$,  GG yields a $2$-approximation to OPT  in $O(m+n\log{m})$ time.
\end{lemma}

\begin{proof} Define intervals $J_i$ as follows:
when capturing a client $p_i$ from a server $t_j$ whose current
radius (prior to capture) is $r_j$, let $J_i=(t_j+r_j,p_i]$ if
$p_i>t_j$, and $J_i=[p_i,t_j-r_j)$ otherwise. Our first trivial
yet crucial observation is that $J_i\cap J_k=\emptyset$ if $i\neq
k$. Also note that the sum of the lengths of the $J_i$ is equal
to the sum of the radii in the GG cover.

Consider now a fixed disk $D$ in OPT, centered at $t_D$, and the
list of intervals  $J_i$ whose $p_i$ is inside $D$. As before, at most one
such $J_i$ extends outward to the right from the right edge of
$D$. If so, call it $J_R$, and define $J_L$ symmetrically. If
$J_R$ exists, it cannot extend more than radius($D$) to the right
of $D$. Let $\lambda=\mathrm{length}(J_R)$. We argue that there is
an interval of length $\lambda$ in $D$, to the right of $t_D$,
which is free of $J_i$'s. It follows that there is at most
radius($D$) worth of segments to the right of $t_D$. Of course,
this is also true if $J_R$ does not exist. By symmetry, there is
also radius($D$) worth of segments to the left of $t_D$, whether
$J_L$ exists or not, yielding the claimed $2$-approximation.

Assume $J_R$ exists. Then the algorithm successively extends $J_R$
by growth to the left up to some maximum point (possibly stopping
right at $p_R$). Since the growth could have been induced by
clients to the right of $J_R$, that maximum point is not
necessarily a client. There is, however, some client inside $D$
that is captured last in this process. This client $p_i$ (possibly
$p_R$) cannot be within $\lambda$ of $t_D$, since otherwise it
would have been captured prior to the construction of~$J_R$.

If there is no client between $t_D$ and $p_i$ we are done, since
then there could be no interval $J_k$ in between. Thus consider
the client $p_{i-1}$ just to the left of $p_i$. Suppose
$d(p_{i-1}, p_i) \geq \lambda$.  Then, if $p_{i-1}$ is eventually
captured from the left, we would have the region between $p_{i-1}$
and $p_i$ free of $J_k$'s and be done. On the other hand, if
$p_{i-1}$ is captured from the right, it must be captured by a
server between $p_{i-1}$ and $p_i$, and that server is at least
$\lambda$ to the left of $p_i$ since otherwise $p_i$ would be
captured by that server prior to $p_R$. This leaves the distance
from the server to $p_i$ free of $J_k$'s.

Hence the only case of concern is if $d(p_{i-1}, p_i) < \lambda$.
Clearly $p_{i-1}$ must not have been captured at the time when
$p_R$ is captured since otherwise $p_i$ would have been captured
before $p_R$, contradicting the assumption that $p_i$ is captured
by growth leftward from $p_R$. Similarly, there cannot be a server
between $p_{i-1}$ and $p_i$, since otherwise both $p_{i-1}$ and
$p_i$ would be captured before $p_R$. Together with the definition
of $p_i$, this implies that $p_{i-1}$ is captured from the left.
Therefore, to the left of $p_{i-1}$, there must be one or more
intervals $\{J_{l_i}\}$ whose length is at least $\lambda$ that
are constructed before $p_{i-1}$ is captured. Similarly, to the
right of $p_i$, there must be some one or more intervals
$\{J_{r_j}\}$ whose length is at least $\lambda$, constructed
before $p_i$ is captured. However, either the last $J_{l_i}$ is
placed before the last $J_{r_j}$ or vice versa.  In the first
case, there are no $\lambda$ length obstructions left in the
left-hand subproblem, so $p_{i-1}$ will be covered, and with
$\lambda$ length obstructions remaining in the right subproblem,
$p_i$ will be captured by growth rightward. The second case is
symmetrical to the first. In either case we have a contradiction.
\end{proof}

To see that the factor $2$ is tight, just consider servers at
$-2+\epsilon, 0$ and $2-\epsilon$ and clients at $-1$ and $1$.

\subsection{Hardness of the two-dimensional discrete problem with superlinear cost}

In 2D, we sketch an NP-hardness proof, for
any $\alpha>1$. This strengthens the NP-hardness proof
of\cite{ESA05}, which only works in the case
$\alpha\geq 2$.

\begin{theorem}
\label{thm:hard}
For any a fixed $\alpha>1$, let the cost function of a circle
of radius $r$ be $f(r)=r^\alpha$. Then it is NP-hard to decide
whether a discrete set of $n$ clients in the plane,
and a discrete set of $m$ potential transmission points allow
a cheap set of circles that covers all demand points.
\end{theorem}

\begin{proofsketch}
Let $I$ be an instance of {\sc Planar 3Sat}, and let $G_I$
be the corresponding variable-clause incidence graph.
After choosing a suitable layout of this planar graph, resulting
in integer variables with coordinates bounded by a polynomial
in the size of $G_I$ for all vertices and edges,
we replace each the vertex representing any particular variable
by a closed loop, using the basic idea shown in the left
of Figure~\ref{fig:clause}; this allows two fundamentally
different ways of covering those points cheaply (using the ``odd''
or the ``even'' circles), representing
the two truth assignments. For each
edge from a vertex to a variable, we attach a similar chain
of points that connects the variable loop to the clause gadget;
the parity of covering a variable loop necessarily assigns
a parity to all incident chains.
Note that choosing sufficiently fine chains guarantees
that no large circles can be used, as the overall weight
of all circles in a cheap solution will be less than 1.
(It is straightforward to see that for any fixed
$\alpha>1$, this can be achieved by choosing coordinates that are
polynomial in the size of $G_I$, with the exponent being $O(1/(\alpha-1))$.)

\begin{figure}[htbp]
\centering{{\includegraphics[width=0.59\columnwidth]{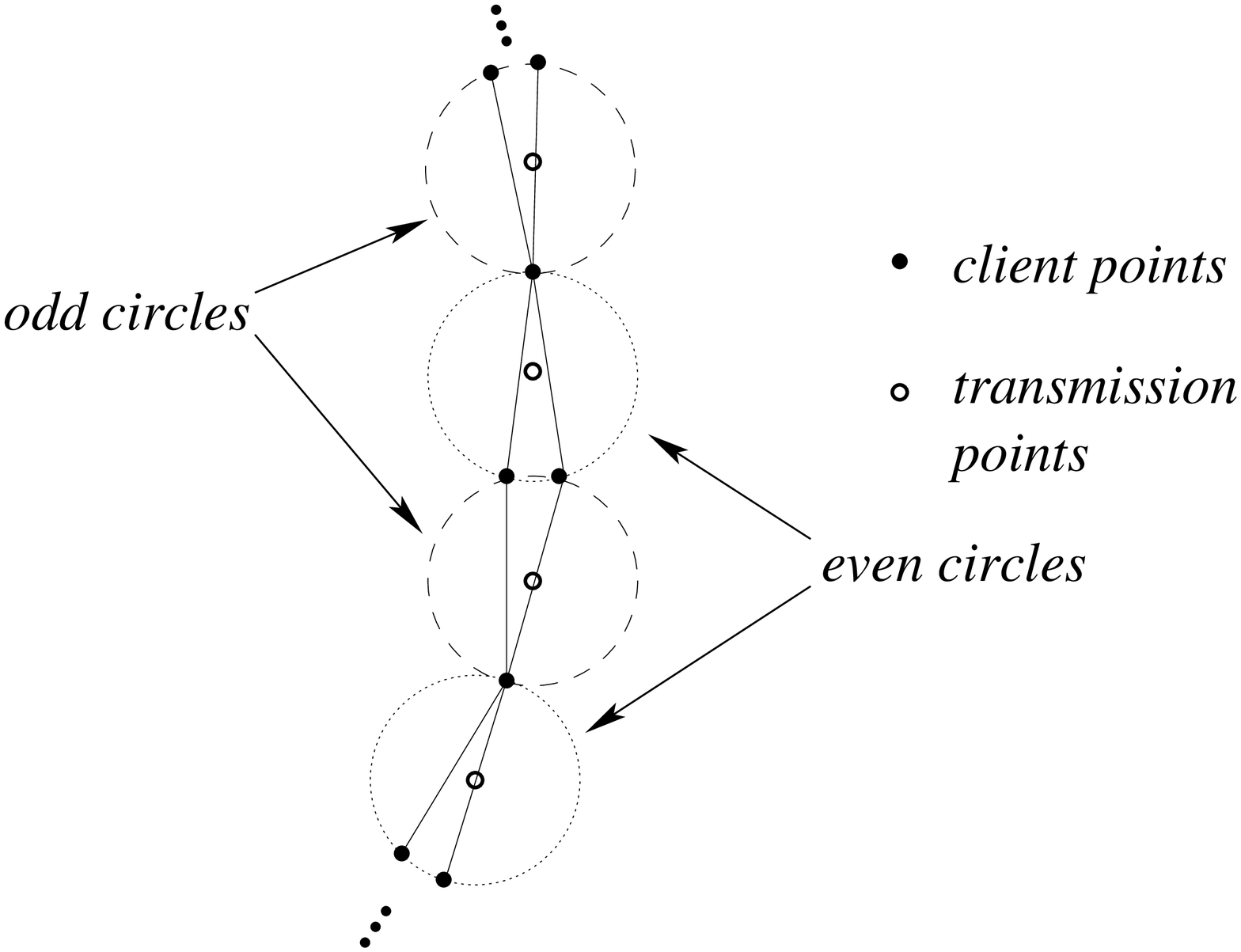}}
{\includegraphics[width=0.39\columnwidth]{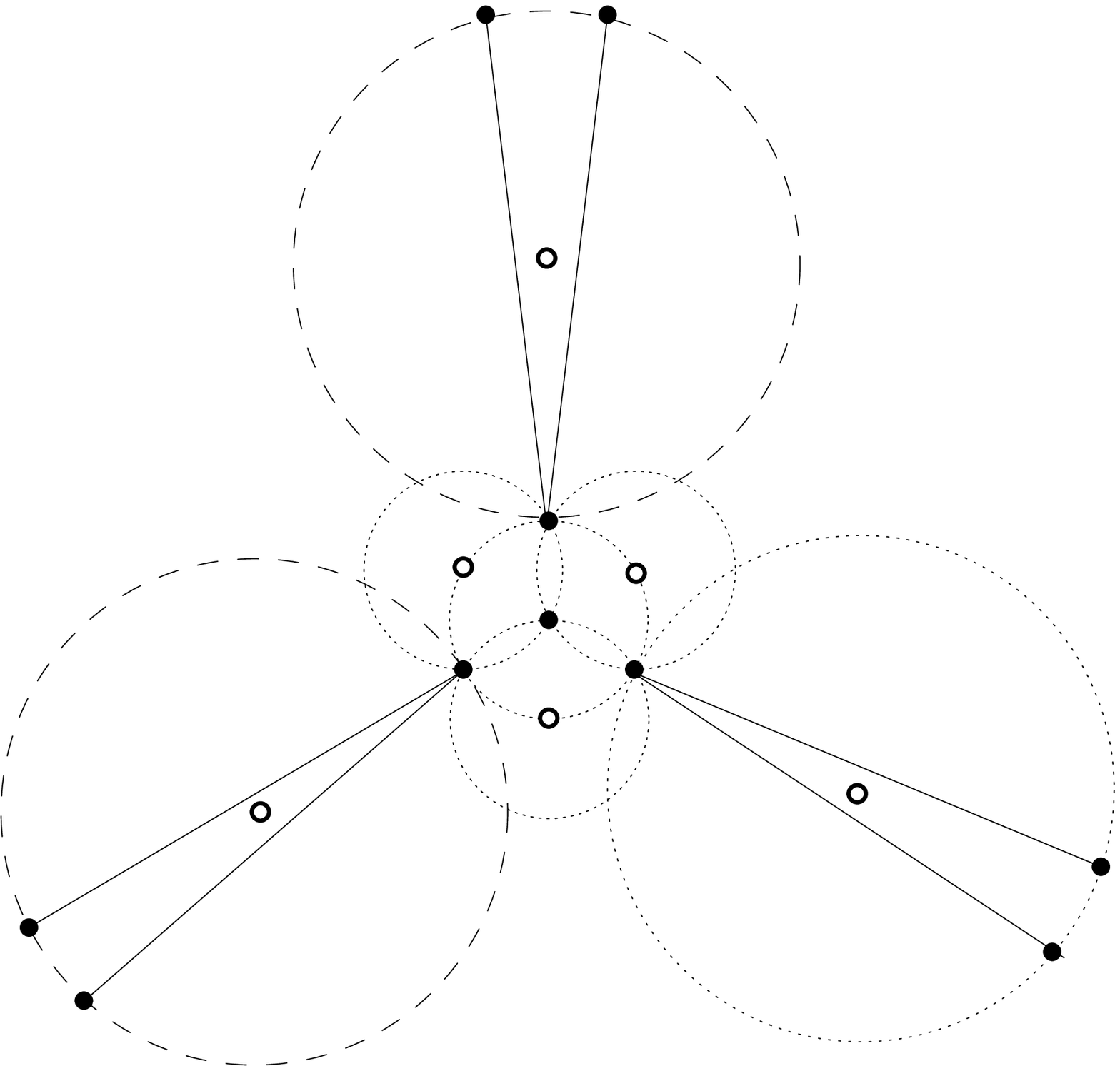}}}
\caption{
  (Left) The switch structure of a variable gadget. Note how there are two
  fundamentally different ways to cover all points cheaply.
  (Right) The structure of a clause gadget. One small circle is needed for
  picking up the client point at the center of the gadget.}
\label{fig:clause}
\end{figure}

For the clauses choose a hexagonal arrangement as shown in the right of
Figure~\ref{fig:clause}: There is one central point that must
be covered somehow; again, $\alpha>1$ guarantees that it is cheaper
to do this from a nearby transmission point, rather than increasing
the size of a circle belonging to a chain gadget.

Now it is straightforward to see that there is a cheap cover, using
only the forced circles, iff the truth assignment corresponding
to the covering of variabe loops assures that each clause has at least
one satisfying variable.
\qed
\end{proofsketch}

%%%%%%%%%%%%%%%%%%%%%%%%%%%%%%%%%%%%%%%%%%%%%%%%%%%%%%%%%%%%%%%%%%%%%%%
\section{Scenario (2): Server Locations\\ Restricted to a Line}
\label{sec:scenario1}
%%%%%%%%%%%%%%%%%%%%%%%%%%%%%%%%%%%%%%%%%%%%%%%%%%%%%%%%%%%%%%%%%%%%%%%

\subsection{Servers along a fixed horizontal line}

\subsubsection{Exact solutions}
\label{sec:1b-exact}

Suppose that the servers are required to lie on a fixed horizontal
line, which we take without loss of generality to be the $x$-axis.
Such a restriction could arise naturally (e.g., the servers must be
connected to a power line, must lie on a highway, or in the main corridor
in a building). In addition, this case must be solved first before
attempting to solve
the more general problem---along a polygonal curve.

In this section, we describe
dynamic programming algorithms to compute a set of server points
of minimum total cost.  For notational convenience, we assume that
the clients $Y$ are indexed in left-to-right order.  Without loss
of generality, we also assume that all the clients lie on or above
the $x$-axis, and that no two clients have the same
$x$-coordinate.  (If a client $p_i$ lies directly above another
client $p_j$, then any circle enclosing $p_i$ also encloses $p_j$,
so we can remove $p_j$ from $Y$ without changing the optimal
cover.)

Let us call a circle $C$ \emph{pinned} if it is the leftmost
smallest axis-centered circle enclosing some fixed subset of
clients.  Equivalently, a circle is pinned if it is the leftmost
smallest circle passing through a chosen client or a chosen pair
of clients.  Under any $L_p$ metric, there are at most $O(n^2)$
pinned circles.  As long as the cost function $f$ is
non-decreasing, there is a minimum-cost cover consisting entirely
of pinned circles.

\mypara{Linear Cost.}
If the cost function $f$ is linear (or sublinear), we easily
observe that the circles in any optimum solution must have
disjoint interiors.  (If two axis-centered circles of radius $r_i$
and $r_j$ intersect, they lie in a larger axis-centered circle of
radius at most $r_i+r_j$.)  In this case, we can give a
straightforward dynamic programming algorithm that computes the
optimum solution under any $L_p$ metric.

The algorithm given in Figure~\ref{2alg} (left)
finds the minimum-cost cover by disjoint
pinned circles, where distance is measured using any $L_p$ metric.
We call the rightmost point enclosed by any pinned circle $C$ the
\emph{owner} of~$C$.

If we use brute force to compute the extreme points enclosed by
each pinned circle and to test whether any points lie directly
above a pinned circle, this algorithm runs in $O(n^3)$ time.  With
some more work, however, we can improve the running time by nearly
a linear factor.

This improvement is easiest in the $L_\infty$ metric, in which
circles are axis-aligned squares.  Each point $p_i$ is the owner
of exactly $i$ pinned squares: the unique axis-centered square
with $p_i$ in the upper right corner, and for each point $p_j$ to
the left of $p_i$, the leftmost smallest axis-centered square with
$p_i$ and $p_j$ on its boundary.  We can easily compute all these
squares, as well as the leftmost point enclosed by each one, in
$O(i\log i)$ time.  (To simplify the algorithm, we can actually
ignore any pinned square whose owner does not lie on its right
edge.)  If we preprocess $P$ into a priority search tree in
$O(n\log n)$ time, we can test in $O(\log n)$ time whether any
client lies directly above a horizontal line.  The overall running
time is now $O(n^2\log n)$.

For any other $L_p$ metric, we can compute the extreme points
enclosed by all $O(n^2)$ pinned circles in $O(n^2)$ time using the
following duality transformation.  If $C$ is a circle centered at
$(x,0)$ with radius $r$, let $C^*$ be the point $(x,r)$.  For each
client $p_i$, let $p_i^* = \{ C^* | \mbox{$C$ is centered on the $x$-axis and } p_i \in C \}$, and let $Y^* =
\{p_i^*\mid p_i\in Y\}$.  We easily verify that each set $p_i^*$
is an infinite $x$-monotone curve.  (Specifically, in the
Euclidean metric, the dual curves are hyperbolas with asymptotes
of slope $\pm 1$.)  Moreover, any two dual curves $p_i^*$ and
$p_j^*$ intersect exactly once; i.e., $Y^*$ is a set of
pseudo-lines. Thus, we can compute the arrangement of
$Y^*$ in $O(n^2)$ time.  For each pinned circle $C$, the dual
point $C^*$ is either one of the clients $p_i$ or a vertex of the
arrangement of dual curves $Y^*$.  A circle $C$ encloses a client
$p_i$ if and only if the dual point $C^*$ lies on or above the
dual curve $p_i^*$.  After we compute the dual arrangement, it is
straightforward to compute the leftmost and rightmost dual curves
below every vertex in $O(n^2)$ time by depth-first search.

Finally, to test efficiently whether any points lie directly above
an axis-centered ($L_p$) circle, we can use the following
two-level data structure.  The first level is a binary search tree
over the $x$-coordinates of~$Y$.  Each internal node $v$ in this
tree corresponds to a canonical vertical slab $S_v$ containing a
subset $p_v$ of the clients.  For each node $v$, we partition the
$x$-axis into intervals by intersecting it with the furthest-point
Voronoi diagram of $p_v$, in $O(\abs{p_v}\log \abs{p_v})$ time.
To test whether any points lie above a circle, we first find a set
of $O(\log n)$ disjoint canonical slabs that exactly cover the
circle, and then for each slab $S_v$ in this set, we find the
furthest neighbor in $p_v$ of the center of the circle by binary
search.  The region above the circle is empty if and only if all
$O(\log n)$ furthest neighbors are inside the circle.  Finally, we
can reduce the overall cost of the query from $O(\log^2 n)$ to
$O(\log n)$ using fractional cascading.  The total preprocessing
time is $O(n\log^2 n)$.

\begin{theorem}\label{thm:DPlinear}
Given $n$ clients in the plane, we can compute in $O(n^2\log n)$
time a covering by circles (in any fixed $L_p$ metric) centered on
the $x$-axis, such that the sum of the radii is minimized.
\end{theorem}

\begin{figure*}[t]
\centerline{\small\begin{algorithm}
$\underline{\textsc{MinSumOfRadiusCircleCover}(Y):}$\+
\\  for every pinned circle $C$\+
\\      find the leftmost and rightmost points enclosed by $C$\-
\\  $Cost[0] \gets 0$
\\  for $i \gets 1$ to $n$ \+
\\      $Cost[i] \gets \infty$
\\      for each pinned circle $C$ owned by $p_i$\+
\\      if no points in $P$ lie directly above $C$\+
\\          $p_j \gets$ leftmost point enclosed by $C$
\\          $Cost[i] \gets \min\{Cost[i],\, Cost[j-1] + radius(C))\}$\-\-\-
\\  return $Cost[n]$
\end{algorithm}\hfill
\small\begin{algorithm}
$\underline{\textsc{MinSuperlinearCostCircleCover}(Y,f):}$\+
\\  sort the pinned circles from left to right by their centers
\\  $Cost[0] \gets 0$
\\  for $j\gets 1$ to $p+1$\+
\\       $Cost[j] \gets \infty$
\\       for $i\gets 1$ to $j-1$\+
\\         if $C_i$ and $C_j$ exclude each other's apices
\\         and $B(C_i, C_j)$ is empty\+
\\             $Cost[j] \gets \min\{Cost[j],\, Cost[i] + f(radius(C_i)))\}$\-\-\-
\\  return $Cost[p+1]$
\\ \quad
\end{algorithm}}
\caption{The dynamic programming algorithm: Left: linear cost; Right: superlinear cost function.}
\label{2alg}
\end{figure*}

\mypara{Superlinear Cost.}
A similar dynamic programming algorithm computes the optimal
covering under any superlinear (in fact, any
\emph{non-decreasing}) cost function~$f$.  As in the previous
section, our algorithm works for any $L_p$ metric.  For the
moment, we will assume that $p$ is finite.

\begin{figure}[ht]
\centering\footnotesize\sf
\begin{tabular}{c@{\qquad}c@{\qquad}c}
    \scalebox{0.70}{\includegraphics[height=1.1in]{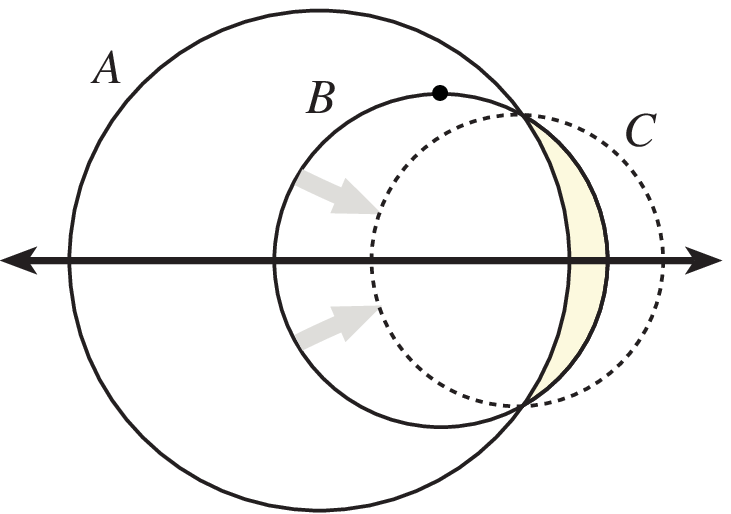}} &
    \scalebox{0.70}{\includegraphics[height=1in]{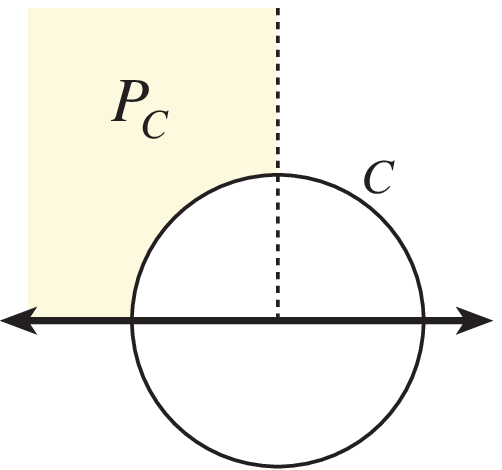}} &
    \scalebox{0.70}{\includegraphics[height=1in]{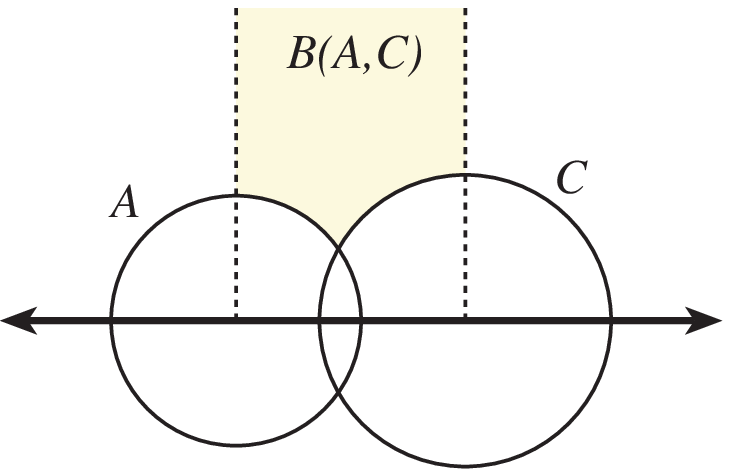}} \\
    (a) & (b) & (c)
\end{tabular}
\caption{(a) The apex of each circle in the optimal cover lies
outside the other circles.  (b) The points $Y_C$ lie in the shaded
region. (c) If $A$ and $C$ are adjacent circles in the optimal
covering, the shaded region $B(A,C)$ is empty.} \label{F:overlap}
\end{figure}

Although two circles in the optimal cover need not be disjoint,
they cannot overlap too much.  Clearly, no two circles in the
optimal cover are nested, since the smaller circle would be
redundant.  Moreover, the highest point (or \emph{apex}) of any
circle in the optimal cover must lie outside all the other
circles.  If one circle~$A$ contains the apex of a smaller circle
$B$, then the lune $B\setminus A$ is completely contained in an
even smaller circle~$C$ whose apex is the highest point in the
lune; it follows that $A$ and $B$ cannot both be in the optimal
cover.  See Figure \ref{F:overlap}(a).

To compute the optimal cover of $Y$, it suffices to consider
subproblems of the following form.  For each pinned circle $C$,
let $Y_C$ denote the set of clients outside $C$ and to the left of
its center; see Figure \ref{F:overlap}(b).  Then for each pinned
circle $C$, we have $cost(Y_C) = \min_{A} (f(radius(A)) +
cost(Y_A))$, where the minimum is taken over all pinned circles
$A$ satisfying the following conditions: (1) The center of $A$ is
left of the center of $C$; (2) the apex of $A$ is outside $C$; (3)
the apex of $C$ is outside $A$; and (4) $A$ encloses every point
in $Y_C\setminus Y_A$.  The last condition is equivalent to there
being no clients inside the region $B(A,C)$ bounded by the
$x$-axis, the circles $A$ and $C$, and vertical lines through the
apices of $A$ and $C$; see Figure \ref{F:overlap}(c).

Our dynamic programming algorithm (Figure~\ref{2alg} (right))
considers the pinned circles
$C_1, C_2, \dots, C_p$ in left to right order by their centers;
that is, the center of $C_i$ is left of the center of $C_j$
whenever $i<j$.  To simplify notation, let $Y_i = Y_{C_i}$.  For
convenience, we add two circles $C_0$ and $C_{p+1}$ of
radius zero, centered far to the left and right of $Y$,
respectively, so that $Y_0= \varnothing$ and $Y_{p+1} = Y$.

Implementing everything using brute force, we obtain a running
time of $O(n^5)$.  However, we can improve the running time to
$O(n^4\log n)$ using the two-level data structure described in the
previous section, together with a priority search tree.  The
region $B(C_i, C_j)$ can be partitioned into two or three
three-sided regions, each bounded by two vertical lines and either
a circular arc or the $x$-axis.  We can test each three-sided
region for emptiness in $O(\log n)$ time.

\begin{theorem}\label{thm:DPsuperlinear}
Let $f:\R_+\to\R$ be a fixed non-decreasing cost function.  Given
$n$ clients in the plane, we can compute in $O(n^4\log n)$ time a
covering by circles (in any fixed $L_p$ metric) centered on the
$x$-axis, such that the sum of the costs of the circles is
minimized.
\end{theorem}

The algorithm is essentially unchanged in the $L_\infty$ metric,
except now we define the apex of a square to be its upper right
corner.  It is easy to show that there is an optimal square cover
in which  no square contains the apex of any other square.
Equivalently, we can assume without loss of generality that if two
squares in the optimal cover overlap, the larger square is on the
left.  To compute the optimal cover, it suffices to consider
subsets $Y_C$ of points either directly above or to the right of
each pinned square $C$.  For any two squares $A$ and~$C$, the
region $B(A,C)$ is now either a three-sided rectangle or the union
of two three-sided rectangles, so we can use a simple priority
search tree instead of our two-level data structure to test
whether $B(A,C)$ is empty in $O(\log n)$ time.

However, one further observation does improve the running time by
a linear factor: Without loss of generality, the rightmost box in
the optimal cover of $Y_C$ has the rightmost point of $Y_C$ on its
right edge.  Thus, there are at most $n$ candidate boxes $C_i$ to
test in the inner loop; we can easily enumerate these candidates
in $O(n)$ time.

\begin{theorem}
Let $f:\R_+\to\R$ be a fixed non-decreasing cost function.  Given
$n$ clients in the plane, we can compute in $O(n^3\log n)$ time a
covering by axis-aligned squares centered on the $x$-axis, such
that the sum of the costs of the squares is minimized.
\end{theorem}

\subsubsection{Fast and simple solutions}
\label{sec:1b-cheap}

In this section we describe simple and inexpensive algorithms that
achieve constant factor approximations for finding a minimum-cost
cover with disks centered along a fixed horizontal line $L$, using any $L_p$ metric.
The main idea for the proofs of this section is to associate with
a given disk $D$ in OPT, a set of disks in the approximate
solution and argue that the set of associated disks cannot be more
than a given constant factor cover of $D$, in terms of cumulative
edge length, cumulative area, and so forth.

As in the previous section, the case of $L_\infty$ metric is the
easiest to handle.  By equivalence of all the $L_p$ metrics,
constant-factor $c$-approximations for squares will extend to
constant-factor $c'$-approximations for $L_p$ disks.

\smallskip
\noindent
\textbf{Square Greedy Cover Algorithm (SG):} Process the client points in
order of decreasing distance from the line $L$.  Find the farthest
point $p_1$ from $L$; cover $p_1$ with a square $S_1$ exactly of
the same height as $p_1$ centered at the projection of $p_1$ on
$L$. Remove all points covered by $S_1$ from further consideration and
recurse, finding the next farthest point from $L$ and so forth. In
the case where two points are precisely the same distance from
$L$, break ties arbitrarily.

Obviously, SG computes a valid covering of $Y$ by construction.We  begin the analysis with  a simple observation.

\begin{lemma} \label{overlay_lemma} In the SG covering, any point in the
plane (not necessarily a client) cannot be covered by more than two
boxes.
\end{lemma}

\begin{proof} Suppose $S_i$ and $S_j$ are two squares placed during
the running of SG and that $i < j$ so that $S_i$ was
placed before $S_j$.  Then $S_i$ cannot contain the center point
of $S_j$ since then $S_j$ would not have had the opportunity to be
placed, and similarly $S_j$ cannot contain the center point of
$S_i$.  Now consider a point $p \in S_i \cap S_j$.  If $p$ were
covered by a third square $S_k$ then either one of $\{S_i, S_j\}$
would contain the center of $S_k$, or $S_k$ would contain the
center of one of $\{S_i, S_j\}$, neither of which is possible.
\end{proof}

\begin{theorem} \label{SG_approx_lemma} Given a set $Y$ of $n$ clients in the plane and any $\alpha\geq 1$,
SG computes in time $O(n \log n)$ a covering of $Y$ by axis-aligned squares
centered on the $x$-axis whose cost is at most three times the optimal.
\end{theorem}

\begin{proof} Let $Y=\{p_1,\ldots,p_n\}$ and consider a
square $S$ in OPT.  We consider those squares $\{S_{i_j}\}$
selected by SG corresponding to points $\{p_{i_j} :
p_{i_j} \in S\}$, see Figure \ref{fig:protrusion_arg},
\begin{figure} [h]
\begin{center}
    \epsfxsize=10cm
    \scalebox{0.70}{\epsfbox{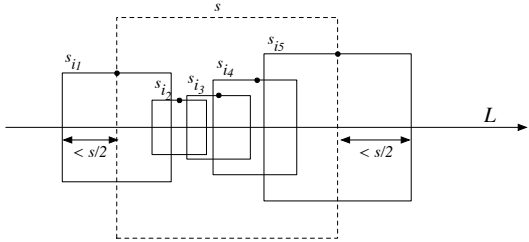}}
%\centerline{\scalebox{0.20}{\includegraphics{images/simple_protrusion_arg2.eps}}}
\caption{Squares of the SG algorithm inside a square
of the optimal solution.}\label{fig:protrusion_arg}
\end{center}
\end{figure}
and argue that these squares cannot have more than three times the
total edge length of $S$.  The same will then follow for all of
SG and all of OPT. The argument, without modification,
covers the case of cost measured in terms of the sum of edge
length raised to an arbitrary positive exponent $\alpha \ge 1$.

Arguing as in Lemma \ref{overlay_lemma} it is easy to see that at
most two boxes $S_{i_j}$ associated with points $p_{i_j} \in S$
processed by SG actually protrude outside of $S$, one
on the left and one on the right.  Denote by $r$ the total
horizontal length of these protruding parts of squares, then $r
\leq s$, the side length of $S$, since the side length of each
protruding square is at most $s$ and at most half of each square
is protruding.

Because of Lemma \ref{overlay_lemma} the total horizontal length
of all nonprotruding parts of the squares $S_{i_j}$ is at most
$2s$, consequently all points covered by $S$ in OPT are covered by
a set of squares $S_{i_j}$ in SG whose total
(horizontal) edge length $\sum_j s_{i_j}$ is at most $3s$.

For exponents $\alpha > 1$ observe that $\sum_j s_{i_j} \le 3s$
and $0 \le s_{i_j} \le s$ for all $j$ implies that $\sum_j
{s_{i_j}}^\alpha \le 3s^\alpha$.

To analyze the running time of the algorithm we need some more
details about the data structures used: Initially, sort the points
by $x$-coordinate and separately by distance from the line $L$ in
time $O(n \log{n})$ and process the points in order of decreasing
distance from $L$. As the point $p_i$ at distance $d_i$ from $L$
is processed, we throw away points which are within horizontal
distance $d_i$ from $p_i$. This takes time $O(\log{n} + k_i)$ time
where $k_i$ is the number of points within $d_i$ from $p_i$. Since
we do this up to $n$ times with $k_1 + \cdots + k_l = n$ the total
running time is $O(n \log{n})$.
\end{proof}

For the linear cost function, it is easy to modify the SG algorithm to get a
$2$-approximation algorithm.

\smallskip
\noindent
\textbf{Square Greedy with Growth Algorithm (SGG):}
Process the points as in SG.
However, if capturing a point $p_i$ by a square $S_i$ would result
in an overlap with already existing square $S_j$ then, rather
than placing $S_i$, grow $S_j$ just enough to capture $p_i$,
keeping the vertical edge furthest from $p_i$ at the same point on
$L$. If placing $S_i$ would overlap two squares,
grow the one
which requires the smallest edge extension.  Break ties arbitrarily.

A proof somewhat similar to that of Lemma~\ref{l:2-approx} shows that:

\begin{theorem} \label{thm:SGG}
Given $n$ clients in the plane, SGG computes in time $O(n \log n)$ a
covering by axis-aligned squares centered on the $x$-axis whose cumulative
edge length is at most twice the optimal.
\end{theorem}

\begin{proof} As we process points $p_i$ using SGG, attribute to
each point $p_i$ a line segment $s_i$ along $L$ as follows.  If
processing $p_i$ resulted in the placement of a square $S_i$
centered at the projection of $p_i$ in $L$ then attribute to $p_i$
the projection on $L$ of a horizontal edge of $S_i$ (Case 1). If,
on the other hand, processing of $p_i$ resulted in the growing of
a prior square $S_j$ to just capture $p_i$, attribute to $p_i$
the projection on $L$ of the portion of the horizontal edge of the
expanded $S_j$ needed to capture $p_i$ (Case 2). (This amount
is at most the distance of $p_i$ to $L$ since otherwise $p_i$ would have
been fallen into case 1.) We must show that the
lengths of the segments is no more than twice the edge lengths of
squares in OPT.

It suffices to show that for any square $S$ in OPT, the segments
$s_i$ associated with points $p_i \in S$ processed by SGG cannot
have total edge length which exceeds twice the edge length $s$ of
$S$.

To see this observe that the sum of the lengths of those $s_i$
lying completely inside $S$ does not exceed $s$ since they are
nonoverlapping. In addition, each of the parts of the at most two
segments protruding from $S$ can have length at most $s/2$, in
case 1 for the same reason as in the SG algorithm, in
case 2 since the total length of the segment is at most $s/2$.

In order to make SGG efficient, we proceed as in
SG. In addition, we maintain a balanced binary search
tree containing the $x$-coordinates of the vertical sides of the
squares already constructed. For each new point $p_i$ to be
processed we locate its $x$-coordinate within this structure to
obtain its neighboring squares and to decide whether case 1 or
case 2 applies. This can be done in time $O(\log n)$ just as
adding a new square in case 1 or updating an existing square in
case 2. Removing points covered by the new or updated square is
done as in SG, so that the total runtime remains $O(n
\log n)$.
\end{proof}

Unlike SG, SGG is not a constant factor approximation for area.
Consider $n$ consecutive points at height $1$ separated one from
the next by distance of $1 + \epsilon$. Processing the points left
to right using SGG covers all points with one square of edge
length $n + (n-1)\epsilon$, and so area $O(n^2)$, while covering
all points with $n$ overlapping squares each of edge length $2$,
uses total area~$4n$.

Finally, extending these results from squares to disks in any $L_p$ metric is not difficult.
Enclosing each square in the algorithm by an $L_p$ disk
leads to an approximation factor $3c^2$ for GG and $2c^2$ for SGG, where $c=p^{\alpha/p}$.
In particular, for $L_2$ disks, this yields a $2\sqrt2$-approximation for $\alpha=1$
and a $4$-approximation for $\alpha=2$.

\subsection{Finding the best axis-parallel line}

When the horizontal line $\ell$ is not given but its orientation
is fixed, we first prove that finding the best line, even for
$\alpha = 1$, is uncomputable, then in this linear case give a
simple approximation, and finally a PTAS.

\subsubsection{A hardness result -- uncomputability by radicals}

Our approach is similar to the approach used by Bajaj on the
unsolvability of the Fermat-Weber problem and other geometric
optimization problems\cite{b-pgans-86,b-adgop-88}.

\begin{theorem}\label{thm:radicals}
Let $c(t)=\sum_i{r_i}$ denote the minimum cost of a cover whose
centers lie on the line of equation $y=t$. There exists a set $Y$
of clients such that, if $t_0$ is the value that minimizes $c(t)$,
then $t_0$ is uncomputable by radicals.
\end{theorem}

The proof proceeds by exhibiting such a point set and showing by
differentiating $c(t)$ that $t_0$ is the root of a polynomial which is
proven not to be solvable by radicals.

The following definitions and facts can be found in a standard
abstract algebra reference; see, for example,
Rotman\cite{rotman-2002}. A polynomial with rational coefficients
is \emph{solvable by radicals} if its roots can be expressed using
rational numbers, the field operations, and taking $k$th roots.
The \emph{splitting field} of a polynomial $f(x)$ over the field
of rationals $\Q$ is the smallest subfield of the complex numbers
containing all of the roots of $f(x)$. The \emph{Galois group} of
a polynomial $f(x)$ with respect to the coefficient field $\Q$ is
the group of automorphisms of the splitting field that leave $\Q$
fixed.  If the Galois group of $f(x)$ over $\Q$ is a symmetric
group on five or more elements, then $f(x)$ is not solvable by
radicals over $\Q$.

Consider the following set of points:$\{(3,4), (-3,-2), (102,2),$ $(98,-2), (200,-2)\}$.
By exhaustive case analysis, we can show that the optimal solution must consist of one circle through the first two points, a second circle through the next two points, and a third circle touching the last point, and the optimal horizontal line must lie in the range $-2\le y \le 2$.
For a given value of $y$ in this range, the cost of the best cover
is
$$
c(y) = \sqrt{2(y-1)^2 + 18} + \sqrt{2y^2 + 8} + (2-y).
$$
Therefore, in order to find the best horizontal line, we must
minimize $c(y)$.  Setting the derivative to zero, we obtain the
equation
$$
c'(y) = \frac{2(y-1)}{\sqrt{2(y-1)^2 + 18}} +\frac{2y }{\sqrt{2y^2
+ 8}} - 1 = 0.
$$
We easily verify that $c''(y)$ is always positive.  The minimum
value $c(y) \approx 8.3327196$ is attained at $y\approx
1.4024709$, which is a root of the following polynomial:
\begin{eqnarray*}
f(y)& = & 1024+512 y-1600y^2+1536y^3-960y^4\\
      &    & +368y^5 -172y^6+28y^7 -7y^8.
\end{eqnarray*}
Using the computational system GAP\cite{GAP4}, we compute that the
Galois group of $f(y)$ is the symmetric group~$S_8$, so the
polynomial is not solvable by radicals.

\subsubsection{Fast and simple constant-factor approximations}

The simple constant factor approximations for a fixed line can be
extended to the case of approximations to the optimal solution on an
arbitrary axis-parallel line with the same constant factors, though
with a multiplicative factor of $O(n^2)$ increase in running time.

\subsubsection{An FPTAS for finding the best horizontal line}

% Joe reworded slightly on 3/26/06:
%
We begin with the case $\alpha = 1$. Let $d$ denote the distance
between the highest and lowest point. Clearly, $d/2 \leq$ OPT $\leq n
d/2$. Partition the horizontal strip of height $d$ that covers the
points into $2n/\epsilon$ horizontal strips, each of height $\delta =
d\epsilon/2n$, using $2n/\epsilon -1$ regularly-spaced horizontal
lines, $\ell_i$. For each line $\ell_i$, we run the exact dynamic
programming algorithm, and keep the best among these solutions.
Consider the line, $\ell^*$, that contains OPT.  We can shift line
$\ell^*$ to the nearest $\ell_i$, while increasing the radius of each
disk of OPT by at most $\delta$, to obtain a covering of the points by
disks centered on some $\ell_i$; the total increase in cost is at most
$\delta n = d\epsilon/2 \leq \epsilon$ OPT. Thus, our algorithm
computes a $(1+\epsilon)$-approximation in time
$O((n^3/\epsilon)\log{n})$.

% Joe rechecked 3/26/06 (it works for an $\alpha>1$, not just integers)
%
In order to generalize this result to the case $\alpha>1$, let us
write PSEUDO-OPT for the lowest cost of a solution on any of the
horizontal lines $\ell_i$, SHIFT for the result of shifting
OPT to the closest of these lines, and $r_1,...,r_m$ for the radii of
the optimal set of disks.  For an arbitrary power $\alpha \ge
1$, we have
\begin{eqnarray*}
  \mbox{PSEUDO-OPT} & \le & \mbox{SHIFT} \le \sum_{i=1}^{m} (r_i+\delta)^\alpha  \\
 & \le & \sum_{i=1}^{m}r_i^\alpha + \delta \alpha \sum_{i=1}^{m} (r_i+\delta)^{\alpha-1}  \\
 & \le & \mbox{OPT}(1 + \delta \alpha 2^{2\alpha-1} n /d).
\end{eqnarray*}
The last line uses $\delta \le d, r_i \le d$ and $\mbox{OPT} \ge
(d/2)^\alpha$. Choosing $\delta = \varepsilon d /(\alpha
2^{2\alpha-1}n)$ gives the desired $(1+\varepsilon)$-approximation.

Together with the results from previous sections we have:

\begin{theorem}
Given $n$ clients in the plane and a fixed $\alpha \geq
1$, there exists an FPTAS for  finding an optimally positioned
horizontal line and a minimum-cost covering by disks centered on
that line. It runs in time $O((n^3/\epsilon) \log n)$ for the
linear cost case ($\alpha=1$) and $O((n^5/\epsilon) \log n)$ for
$\alpha>1$.
\end{theorem}

\subsection{Approximating the best line --\\ any orientation}

Finally, we sketch approximation results for selecting the best
line whose orientation is not given. We give both a constant factor
approximation and a PTAS for the linear cost case ($\alpha=1$).

\subsubsection{Fast and simple constant-factor approximations}

Given a line $\ell$, we say that a set $\mathcal{D}$ of disks $D_1$,\ldots,$D_k$
is \emph{$\ell$-centered} if the centers of every disk $C_j$ in $\mathcal{D}$ belongs to $\ell$.
Recall that the cost of $\mathcal{D}$ is the sum of all its radii.

\begin{lemma}\label{lem:1}
Given $k\geq 1$, a line $\ell$, an $\ell$-centered set $\mathcal{D}$
of $k$ disks that cover $Y$, and any point $p_0$ on $\ell$,
there exist $p'\in Y$ and an $\ell'$-centered set $\mathcal{D}'$ of $k$ disks
that cover $Y$, where $\ell'$ is the line that joins $p_0$ and $p'$,
such that the cost of $\mathcal{D}'$ is at most $2^\alpha$ times the cost of $\mathcal{D}$.
\end{lemma}
\begin{proof}
We will assume without loss of generality that $\ell$ is the $x$-axis, $p_{0}$
is the origin and that no other point in $Y$ lies on the $y$-axis.
The latter restriction can easily be enforced by a small perturbation.
Let the coordinates of $p_i$ be $x_i$ and $y_i$,
and let $m_{i}$ denote the slope $y_{i}/x_{i}$ of the line $\ell_i$
for $1\leq i\leq n$. First, we reorder $Y$ so that $|m_{1}| \leq \dots \leq |m_{n}|$.
In what follows we assume that $x_{1}> 0$ and $y_{1}\geq 0$.  The other
cases can be treated analogously.

For each disk $D_j=D(t_j,r_j)$ in $\mathcal{D}$, we construct
a disk $D'_j$ whose radius is $r'_{j} = 2r_{j}$ and center $t'_j$ is obtained from $t_{j}$ by
rotating it around the origin counterclockwise by an angle $\atan(m_{1})$.
The set $\mathcal{D}'$ of $k$ disks thus defined is $\ell'$-centered,
where $\ell' = \{(x,y) \in \R^{2} \mid
y=m_{1}x\}$ and $p_{1} \in \ell'$.
To see that $\mathcal{D}'$ covers $Y$, simply observe that
$d(t_{j},t'_{j}) \leq r_{j}$ for all $1 \leq j \leq k$ and apply the
triangle inequality: any point in $D_j$ must be at distance
at most $2r_j$ of $t'_j$.
The cost of this new solution is clearly at most $2^\alpha$ times that of $\mathcal{D}$ in the linear cost case.
\end{proof}

By a double appplication of this lemma, first about an arbitrary $p_0$ yielding a point $p'=p_i$, then about
$p_i$ yielding another $p'=p_j$, it is immediate that any $\ell$-centered cover of $Y$ can be
transformed into an $\ell_{i,j}$-centered cover whose cost is increased
at most four-fold, where $\ell_{i,i}$ is the line joining $p_i$ and $p_j$.
By computing (exactly or approximately) the optimal set of disks
for all $O(n^{2})$ lines defined by two different points of $Y$,
we conclude:

\begin{theorem} \label{thm:1}
Given $n$ clients in the plane
and a fixed $\alpha \geq 1$,
  in $O(n^{4} \log n)$ time, we can find a collinear set of disks that cover $P$ at cost at
  most $4^\alpha OPT$, and for $\alpha=1$, in $O(n^{3} \log n)$ time, we can find a collinear set
  of disks that cover $P$ at cost at most $8\sqrt{2}OPT$.
\end{theorem}

\subsubsection{A PTAS for finding the best line with\\ unconstrained orientation}

We now prove that finding the best line with
unconstrained orientation and a minimum-cost covering with disks whose
centers are on that line admits a PTAS.

\begin{theorem}
\label{thm:any-slope}
  Let $Y$ be a set of $n$ clients in the plane that can be covered by an
  optimal collinear set of disks at linear cost $OPT$ (i.e., $\alpha=1$), and $\epsilon>0$.  In
  $O((n^{4}/\epsilon^2) \log n)$ time, we can find a collinear set of disks
  that cover $Y$ at cost at most $(1+\epsilon)OPT$.
\end{theorem}

\begin{proof}
Let $H$ be a strip of minimal width $h$ that contains $Y$.
Using a rotating calipers approach, $H$ can be computed in $O(n \log n)$ time.
If $h=0$, we can conclude that $OPT=0$ and we are done.

Otherwise, we can assume wlog that $H$ is horizontal and that its center line is the $x$-axis.
Let $R$ denote the smallest enclosing axis-parallel rectangle $R$ of $Y$,
$w$ its width, and $h$ its height, Then $h \le w$ and, moreover, $h/2 \leq OPT$.
Let $\ell^*$ be the optimal line.

We now distinguish two cases:

{\bf Case 1. $w \geq 2h$:}
Observe that both vertical sides $v_1,v_2$ of $R$ contain a point of $Y$.
Therefore, $\ell^*$ must have distance at most $OPT$ to $v_1$ and $v_2$.
A straightforward calculation shows that then $\ell^*$ must intersect the lines $\ell_1$ and $\ell_2$
extending $v_1$ and $v_2$ at a distance of at most $4 OPT$ from the $x$-axis.

The idea is now to put points on those parts of $\ell_1$ and $\ell_2$ which are equally spaced at distance
$\delta = \varepsilon OPT / n$.
Then we consider all lines passing through one of these points on $\ell_1$ and one on $\ell_2$.
For each such line we find the optimal covering of $P$ by circles centered on it using the
algorithm of Theorem~\ref{thm:DPlinear},
and give out the best one as an approximation for the optimum.

Observe, that there is one of the lines checked, $\hat{\ell}$, whose intersection points with
$\ell_1$ and $\ell_2$ are at distance at most $\delta/2$ from the ones of $\ell^*$.
Elementary geometric considerations show that to any point
$p$ in $\ell^*$ closest to some point of $P$ there is a point
in $\hat{\ell}$ within distance at most $\delta$. Consequently, to any circle of radius $r$ of the
optimal covering centered on $\ell^*$, there is a circle on $\hat{l}$ of radius $r+\delta$ covering
the same set of points (or more). Thus, $\hat{l}$ has a covering that differs by at most
$ n \delta$ from the optimal one. By the choice of $\delta$ we have a $1+\varepsilon$-approximation
to the optimum.

Observe, that we chose $O(OPT/\delta)=O(n/\varepsilon)$ points on $\ell_1$ and $\ell_2$,
so we are checking $O((n/\varepsilon)^2)$ lines.
For each of them, we apply the algorithm of Theorem~\ref{thm:DPlinear}
which has runtime $O(n^2 \log n)$ yielding a total runtime of $O(n^4 / {\varepsilon^2} \log n)$.

{\bf Case 2. $w < 2h$:}
In this case the optimal line $\ell^*$ can have a steeper slope and even be vertical.
Of course, it must intersect $R$ and we expand $R$ to a cocentric rectangle $R'$ such that
the footpoint of any point in $Y$ on $\ell^*$ must lie inside $R'$.
An easy geometric consideration shows that extending the width of $R$ by $h$ and its height
by $w$ will suffice, so $R'$ is a square of side length $w+h$. Then
we put equally spaced points of distance $\delta = \varepsilon OPT / n$ on the whole boundary of
$R'$, apply the algorithm of Theorem~\ref{thm:DPlinear}
to all lines passing through any two of these points, and return the one giving the smallest
covering as an approximation to the optimum. The same consideration as in the first case shows
that this is indeed a $(1+\varepsilon)$-approximation.
Since the length of the boundary of $R'$ is
$4(w+h) \leq 12h \leq 24 OPT$, we obtain the desired runtime in this case, as well.

For both cases it remains to show how to obtain a suitable value of $\delta$,
since we do not know the value of $OPT$. Since any value below $OPT$ suffices, we simply
run a constant factor $c$ approximation algorithm of Theorem~\ref{thm:DPlinear}
and take $1/c$ times the value it returns instead of $OPT$ in the definition of $\delta$.
\end{proof}

%%%%%%%%%%%%%%%%%%%%%%%%%%%%%%%%%%%%%%%%%%%%%%%%%%%%%%%%%%%%%%%%%%%%%%%
\section{Minimum-Cost Covering Tours}
\label{sec:scenario3}
%%%%%%%%%%%%%%%%%%%%%%%%%%%%%%%%%%%%%%%%%%%%%%%%%%%%%%%%%%%%%%%%%%%%%%%

We now consider the minimum cost
covering tour (MCCT) problem: Given $k\geq 1$ and a set $Y=\{p_1,\ldots,p_n\}$
of $n$ clients, determine
a cover of $Y$ by (at most) $k$ disks centered at $X=\{t_1,\ldots,t_k\}$ with radii $r_j$ and a tour $T$ visiting $X$,
such  that the cost $\mathrm{length}(T)+C\sum r_i^\alpha$ is minimized.
We refer to the tour $T$, together with the disks centered on $X$,
as a \emph{covering tour} of $Y$.
Our results are for the case of linear transmission costs ($\alpha=1$).
We first show a weak hardness result, then characterize the
solution for $C\leq 4$, and finally give a PTAS for a fixed $C>4$.

\subsection{A hardness result}

We prove the NP-hardness of MCCT
where $C$ is also part of the input. Note that this does not prove
the NP-hardness of MCCT where $C$ is a fixed constant, which
is the problem for which we give a PTAS below.
Note also that $C$ appears in the run time exponent of that PTAS,
and so the PTAS no longer runs in polynomial time
if $C$ is not a fixed constant.

\begin{theorem}
\label{thm:hard-MCCT}
MCCT with linear cost is NP-hard if the ratio $C$ is part of the
input.
\end{theorem}

\begin{proofsketch}
We show a reduction from {\sc Hamilton cycle in grid graphs}.
Given a set of $n$ points on a grid, we construct an instance of
MCCT in which each of the given points is a client. We set $C$
to be larger than $2n$. We claim that the grid graph has a
Hamilton cycle if and only if there is a tour $T$ visiting a set
of disk centers with radii $r_i$ whose cost is at most~$n$.

Clearly a Hamilton cycle in the grid graph yields a tour of cost
$n$ with each client contained in a disk of radius 0 centered at
that point.

Conversely, suppose we have a tour whose cost (length plus sum of
radii) is at most $n$. Note that no two clients can be contained
in a single disk, as such a disk must have radius at least 0.5,
and thus its contribution to the cost $C\cdot r_i >2n\cdot 0.5 =n$
contrary to our assumption. Next we want to show that each disk in
an optimal solution is centered at the client it covers. Suppose
this is not the case, there is some client $j$ which is covered by
a disk centered at $c_j\not= j$. Let the distance between client
$j$ and the center of the disk covering it be $d$.  Now consider
an alternate feasible solution in which the tour visits $c_j$ then
$j$ then back to $c_j$, covering $j$ with a disk of radius 0. No
other client is affected by this change, as the disk only covers
point $j$. The cost of the new solution is the cost of the
original (optimal) solution $+2d-Cd$ as we add $2d$ to the length
of the tour, but decrease $C\sum r_i$ by $Cd$. Since $C>2$ the new
solution is better than the original optimal solution, a
contradition.\qed
\end{proofsketch}

\subsection{The case {\Large $C\leq 4$}: The exact solution is\\ a single circle}

\begin{theorem} \label{c=4_lemma} In the plane, with a cost function
of  $\mathrm{length}(T)+C \sum r_i$ and $C \leq 4$, the minimum-cost solution is
to broadcast to all clients from the circumcenter of the client
locations and no tour cost.
\end{theorem}

The proof rests on the following elementary geometry lemma
(whose proof is omitted here).

\begin{lemma} \label{3_point_lemma} For three points $p$, $q$ and $r$ in the plane,
such that the triangle $pqr$ contains its own circumcenter,
the length of a trip from $p$ to $q$ to $r$ and back to $p$ is at
least $4r$ where $r$ is the circumradius of the points.
\end{lemma}

\begin{proofof}{Theorem \ref{c=4_lemma}} Let $r(X)$ and $r(Y)$
denote the minimum radius of a circle enclosing $X$ or $Y$,
respectively.
Let $T$ be a covering tour of $Y$,
$X\subseteq T$ be the set of disk centers and $r_j$ their radii.
Finally, let $r_{\mathrm{max}}=\max_j r_j$.

By the triangle inequality, Lemma \ref{3_point_lemma} implies that
the $\mathrm{length}(T)\geq 4r(X)$.
Since the tour visits all the centers in $X$ and the disks centered at $X$ cover $Y$,
we have $r(Y)\leq r(X)+ r_{\mathrm{max}}$.
By definition, the cost of $T$ is $\mathrm{length}(T) + C\sum_j{r_j}$,
which by the observation above is at least
$4 r(X) +  C\sum_j{r_j} \geq 4r(X) + C r_{\mathrm{max}}$.
The assumption $C\leq 4$ then implies that it be
at least $C(r(X)+ r_{\mathrm{max}})\geq Cr(Y)$, which
is the cost of covering by a single disk with a zero-length tour. \qed
\end{proofof}

\subsection{The case {\Large $C>4$}: A PTAS}

We distinguish between two cases for the choice of transmission
points: they may either be arbitrary points in the plane (selected
by the algorithm) or they may be constrained to lie within a
discrete set ${\cal T}$ of candidate locations.

The constant $C$ specifies the relative weight associated with the two
parts of the cost function -- the length of the tour, and the sum of
the disk radii.  If $C$ is very small ($C\leq 4$), then the solution
is to cover the set $Y$ using a single disk (the minimum enclosing
disk), and a corresponding tour of length 0 (the singleton point that
is the center of the disk).  If $C$ is very large, then the priority
is to minimize the sum of the radii of the $k$ disks.  Thus, the
solution is to compute a covering of $Y$ by $k$ disks that minimizes
the sum of radii (as in\cite{lp-ptasb-05}), and then link the
resulting disk centers with a traveling salesman tour (TSP).  (In the
case that $k\geq n$, the disks in the covering will be of radius 0,
and the problem becomes that of computing a TSP tour on~$Y$.)
Note that our algorithm gives an alternative to the Lev-Tov and Peleg
PTAS\cite{lp-ptasb-05} for coverage alone.

Our algorithm is based on applying the $m$-guillotine
method\cite{m-gsaps-99}, appropriately adapted to take into
account the cost function and coverage constraint.\footnote{The ``$m$''
in this section refers to a parameter, which is $O(1/\eps)$, not
the number of servers.}
We need several definitions; we largely follow the notation of
\cite{m-gsaps-99}.  Let $G=(V,E)$ be an embedding of a connected planar
graph, of total Euclidean edge-length $L$. Let ${\cal D}$
be a set of disks centered at each vertex $v$ of $G$ of radius $r_v$.
We refer to the pair $(G,{\cal D})$ as a {\em covering network} if the
union $\cup_{v\in V} D_v$ of the disks covers the clients $Y$.
We can assume without loss of generality that $G$ is restricted to
the unit square $B$, i.e., $\cup_{e\in E} e\subset int(B)$.

Our algorithm relies on there being a polynomial-size set of
candidate locations for the transmission points that will serve as
the vertices of the covering tour we compute.  In the case that a
set ${\cal T}$ of candidate points is given, this is no issue;
however, in the case that the transmission points are arbitrary,
we appeal to the following grid-rounding lemma (proved in the full paper).

\begin{lemma}
\label{lem:grid} One can perturb any covering network $(G,{\cal
D})$ to have its vertices all at grid points on a regular grid of
spacing $\delta=O(\epsilon \cdot diam(S)/n)$, while increasing the
total cost by at most a factor of $(1+\epsilon)$.
\end{lemma}

An axis-aligned rectangle, $W\subseteq B$, is called a {\em
window}; rectangle $W$ will correspond to a subproblem in a
dynamic programming algorithm.  An axis-parallel line $\ell$ that
intersects $W$ is called a {\em cut}.

For a covering network with edge set $E$ and a set of disks ${\cal
  D}$, we say that $(E,{\cal D})$ satisfies the {\em $m$-guillotine
  property with respect to window $W$} if either (1) all clients
$Y\subset W$ lie within disks of ${\cal D}$ that intersect the
boundary of $W$; or (2) there exists a cut $\ell$ with certain properties
(an \emph{$m$-good cut with respect to $W$})
that splits $W$ into $W_1$ and $W_2$,
and $(E,{\cal D})$ recursively
satisfies the $m$-guillotine property with respect to both $W_1$
and~$W_2$.
Due to the lack of space, we cannot give the full definition of an $m$-good cut
(see the full paper).

The crux of the method is a structural theorem, which shows how to
convert any covering network $(G,{\cal D})$ into another covering
network $(G',{\cal D}')$, such that the new graph $G'$ satisfies the
$m$-guillotine property, and that the total cost of the new instance
$(G',{\cal D}')$ is at most $O((L+CR)/m)$ times greater than the
original instance $(G,{\cal D})$, where $L$ is the total edge length
of $G$ and $R$ the sum of the radii of ${\cal D}$.  The construction
is recursive: at each stage, we show that there exists a cut with
respect to the current window $W$ (which initially is the unit square
$B$), such that we can ``afford'' (by means of a charging scheme) to
add short horizontal/vertical edges in order to satisfy the
$m$-guillotine property, without increasing the total edge length too
much.

We then apply a dynamic programming algorithm,
running in $O(n^{O(m)})$ time,
to compute a minimum-cost covering network
having a prescribed set of properties:
(1) it satisfies the $m$-guillotine property (with respect to $B$),
which is necessary for the dynamic program to  have the claimed efficiency;
(2) its disks cover the clients $Y$; and
(3) its edge set contains an Eulerian subgraph.
This third  condition allows us to extract a tour in the end.
In the proof of the following theorem (see the full paper), we give
the details of the dynamic programming algorithm that yields:

\begin{theorem}
\label{cor:ptas} The min-cost covering tour problem has a PTAS that runs in time $O(n^{O(1/\epsilon)})$.
\end{theorem}

%%%%%%%%%%%%%%%%%%%%%%%%%%%%%%%%%%%%%%%%%%%%%%%%%%%%%%%%%%%%%%%%%%%%%%%
\section*{Acknowledgments}
%%%%%%%%%%%%%%%%%%%%%%%%%%%%%%%%%%%%%%%%%%%%%%%%%%%%%%%%%%%%%%%%%%%%%%%

We thank all of the participants of the McGill-INRIA International
Workshop on Limited Visibility, at the Bellairs Research Institute of
McGill University, where this research was originated.  We
acknowledge valuable conversations with Nancy Amato, Beppe Liotta,
and other workshop participants and heartily  thank the organizers
Sue Whitesides and Hazel Everett for facilitating and enabling a
wonderful working environment.

\balance

\edef\baselinestretch{0.93}

\bibliographystyle{abbrv}
%\bibliography{refs}

\edef\baselinestretch{1.0}
\normalsize

\end{document}